\documentclass[structabstract]{aa}  

\usepackage{graphicx}
\usepackage{txfonts}
\usepackage{natbib}
\bibpunct{(}{)}{;}{a}{}{,} 
\newcommand{\kms}{$\,$km$\,$s$^{-1}$}

\begin{document}

\title{Stokes imaging polarimetry using image restoration at the 
        Swedish 1-m Solar Telescope}  

\author{M.~J. van Noort
  \inst{1}
  \and
  L.~H.~M. Rouppe van der Voort
  \inst{2}
}

\institute{Institute for Solar Physics of the Royal Swedish Academy of
  Sciences, AlbaNova University Center, SE--106\,91 Stockholm,
  Sweden\\
  \email{noort@astro.su.se}
  \and
  Institute of Theoretical Astrophysics, University of Oslo,
  P.O. Box 1029 Blindern, N--0315 Oslo, Norway\\
  \email{rouppe@astro.uio.no}  
  \thanks{also at the Center of Mathematics for Applications,
  University of Oslo, P.O. Box 1053 Blindern, N--0316 Oslo, Norway}
}

\date{Received / Accepted}

\abstract
{}
{We aim to achieve high spatial resolution as well as high
  polarimetric sensitivity, using an earth-based 1m-class solar
  telescope, for the study of magnetic fine structure on the Sun.}
{We use a setup with 3 high-speed, low-noise cameras to construct
  datasets with interleaved polarimetric states, particularly suitable
  for Multi-Object Multi-Frame Blind Deconvolution image
  restorations. We discuss the polarimetric calibration routine as
  well as various potential sources of error in the results.}
{We obtained near diffraction limited images, with a noise level of
  $\approx 10^{-3} I_{cont}$. We confirm that dark-cores have
  a weaker magnetic field and at a lower inclination angle with
  respect to the solar surface than the edges of the penumbral
  filament. We show that the magnetic field strength in
  faculae-striations is significantly lower than in other nearby parts
  of the faculae.}
{}


\titlerunning{Stokes imaging polarimetry using image restoration} 
\authorrunning{van Noort \& Rouppe van der Voort}

\maketitle

\section{Introduction}
\label{sec:intro}

Since the discovery of magnetic features on the Sun
\citep{1908ApJ....28..315H}, 
polarimetry has become an increasingly important diagnostic tool for
solar observations. The primary goals in resolution, required signal
to noise ratio (S/N) and spectral resolution have continually shifted
with the objects of interest, from strong, large-scale fields to
smaller and often weaker structures.

Considerable efforts toward meeting these requirements have been made
in recent times.
Here we mention a few examples of modern instruments at some of the
major solar observing facilities:
%
The Diffraction-Limited Spectro-Polarimeter
\citep[DLSP,][]{2004SPIE.5171..207S},
the Tenerife Infrared Polarimeter \citep[TIP,][]{1999ASPC..183..264M,
  2007ASPC..368..611C},
the Polarimeteric Littrow Spectrograph \citep[POLIS,][]{2005A&A...437.1159B}, 
and the multi-line spectropolarimetric (MTR) observing mode at Themis
\citep{2001ASPC..236....9P}. 
Typical spatial resolution achieved with these instruments is of order
0\farcs5--1\arcsec.

Major progress has been made by the spectro-polarimeter
\citep[SP,][]{2001ASPC..236...33L} on the Solar Optical Telescope
\citep[SOT,][]{2008SoPh..tmp...26S} 
on board the \emph{Hinode}
spacecraft. 
The combination of $\sim$0\farcs3 spatial resolution and
10$^{-3}$--10$^{-4}$ relative sensitivity is setting new standards in
solar polarimetry.
The excellent performance of the image stabilizer of SOT enables the
SP to fully exploit one of the major advantages of space observations:
the absence of seeing.

Conducting sensitive Full Stokes measurements on solar image data from
the ground presents several observational challenges. The problem
poses two basic requirements on the data, that are difficult to
accommodate at the same time. In order to maximize resolution, as short
as possible exposure times are desirable to minimize smearing by
seeing induced image motions, however, to maximize sensitivity, as
long as possible exposure times are desirable in order to minimize the
photon noise in the images. Obviously these two requirements are
mutually exclusive so that a suitable compromise must be made.

Two main approaches currently exist toward addressing one or both of these 
problems. The first approach has been very successful at eliminating
artificial signals resulting from seeing-induced image motions, by means of 
high-frequency modulation of the the polarimetric state, and integrating in all 
polarimetric states for a long time. As the seeing-induced image aberrations 
are identical for all intents and purposes, artifacts resulting from 
image-motions are essentially eliminated. 
Using this method, a relative sensitivity of $10^{-5}$ has been
achieved by the Z{\"u}rich Imaging Stokes Polarimeter II
\citep[ZIMPOL,][]{1997A&A...328..381G}.
Such extreme levels of sensitivity can currently only be achieved at the cost of
losing a great deal of spatial resolution.

Another approach is through the removal of seeing-induced image 
aberrations by numerical means. This approach promises to address both problems at
once, but usually requires frequent CCD readouts to construct an image stack to 
analyze. Consequently, it is of particular importance to ensure that the 
photon noise exceeds the readout noise of the CCD, in order to achieve a good S/N 
in the end result.

In a recent work, \citet{bellogonzalez08speckle}, achieved a
polarimetric sensitivity of $10^{-3}$
at a spatial resolution of $\approx$0\farcs3 using a combination of
adaptive optics and speckle reconstruction.
Their observing method elaborates on the pioneering work of
\citet{1992A&A...261..321K} at the 
former Swedish Vacuum Solar Telescope, where they introduced the use
of a narrow-band channel in combination with a wide-band channel.  The
high light levels in the wide-band channel enabled them to
successfully perform speckle restorations, after which the seeing
information acquired from the wideband restorations was used to
restore the narrowband images.

The Multi-Object Multi-Frame Blind Deconvolution (MOMFBD)
post-processing method used by \cite{vanNoort05MOMFBD}, like the speckle
method used by \citet{1992A&A...261..321K}, combines the signal
accumulated in a large number of short-exposure images using an image
restoration process that separates the image information from the
seeing-induced image blurring and distortions. Although coming at a
significantly increased numerical cost compared to speckle reconstruction, 
some of the advantages of the
MOMFBD method are that some fixed (instrumental) aberrations can be
corrected using a phase-diversity channel, the process uses all
channels for wavefront sensing, resulting in better separation of
image information and seeing distortions, and the requirements on the
image data are more relaxed regarding the statistical independence of
individual exposures and on the seeing statistics in general, so that
it can be easily applied to very high cadence data, recorded using an 
adaptive optics system.

\section{Full Stokes measurements at the SST}
\label{sec:fullstokes}

In order to take accurate polarimetric data, a reliable polarimeter is
needed.  Although strictly speaking the whole telescope is part of the
polarimeter, it makes sense to separate the the vacuum part of the
telescope from the rest of the optical setup and model it separately.

The reason for this is that the design of the %
Swedish 1-m Solar Telescope \citep[SST,][]{scharmer2003SST} %
includes vacuum windows and
large angle of incidence mirrors, so that the instrumental
polarization is not only considerable, at $\approx$10\% in both Stokes
Q and U and around 2\% in Stokes V, it also depends on the mirror
angles and thus on the observing position in the sky and is therefore
time dependent.

As determination of the telescope matrix for all possible positions in
the sky is not practically achievable (although not impossible), it
makes sense to model it and calibrate the model parameters
instead. For this, however, an accurate polarimeter is needed first.

\subsection{Polarimeter}
The polarimeter at the SST is similar to that used by \citet{1999ASPC..184...33H} 
and consists of two Liquid Crystal Variable
Retarders (LCVR), positioned in the beam, the first aligned with a
vertical analyzer polarization filter, which is part of the SOUP
filter, and the second at a 45\degr\ angle with respect to the
first. By varying the retardance of each LCVR and filtering the resulting 
"modulated" light through an analyzer Linear Polarizer (LP), the polarimeter can be
configured to transmit only certain (linearly independent) combinations of the
input Stokes parameters. The conversion coefficients
from Stokes parameters to transmitted intensity are the elements of the 
modulation matrix $M$, the inverse of which, the demodulation matrix, 
must be known to retrieve the input Stokes parameters.

The choice for LCVRs over the more commonly used Ferro-Electric LCs,
is motivated by the greater flexibility in choosing the modulation
scheme and effortless adaptation to changes in the observing
wavelength. The price to pay for this flexibility is a "slow" response to changes
in modulation value ($\sim$10~ms or more) when an increase in
retardance is required, which presents a problem when combined with
high-speed cameras, as well as stringent stability requirements on the
driving voltage, especially at the high end of the retardance range of
the LCVRs, where the sensitivity to voltage changes is largest.

The speed restrictions of the LCVRs can be effectively dealt with
using an "overdrive" approach: when an increase in retardance is
needed, the driving voltage amplitude is set to 0, in order to
maximize the rate of change. When the desired retardance value has
been reached (or passed), the voltage appropriate to the target
retardance value is applied, upon which the LCVR has only a small,
negative, retardance change to cover, which can be accomplished more
than an order of magnitude faster than for an equal positive
change. In combination with optimizing the state order of the
modulation scheme, the tuning time of the LCVRs can be kept below
10~ms at all times, a requirement imposed by the readout time of the
cameras used.

Since the retardance value of Liquid Crystals is also temperature
sensitive, the ability to continuously vary the retardance value is an
advantage, as it can be used to correct the retardance based on the
measured temperature, so that complications related to thermal
stabilization can be avoided.

The current implementation, however, still makes use of a simple
thermal stabilization circuit, which, although it improves
considerably upon the unstabilized situation, is not able to cope with
sudden changes in the environmental conditions, such as those arising
from changes in ventilation, etc.  Therefore, although the polarimeter
is currently operational, its properties are slowly varying over time
and need to be recalibrated as frequently as possible. Once
the stability issues are properly addressed, this should not be
needed, as long as the optical setup remains the same.

For most of the data presented here, the polarimeter could be
calibrated within a few hours of taking the data, but uncorrected
variations in the measured Stokes parameters at the level of
approximately one percent usually remain.

\subsection{Polarimeter Calibration}
To calibrate the polarimeter, a sufficiently large number of linearly
independent Stokes vectors needs to be generated. This is done using a
combination of a Linear Polarizer (LP) and a Quarter Wave Plate (QWP):
a 90\degr\ linear retarder. The LP is positioned at several fixed
angles, after which the retarder is rotated through 360+ degrees, so
that some of the linearly polarized light is converted to other Stokes
parameters according to
\begin{equation}
\tiny
\begin{array}{c}
S(\varphi,\alpha)=\\
\frac{1}{2}
\left(\begin{array}{c c c c}
1&0&0&0\\
0&\cos^22\alpha+\sin^2 2\alpha\cos\delta&\cos2\alpha\sin2\alpha(1-\cos\delta)&-\sin2\alpha\sin\delta\\
0&\cos2\alpha\sin2\alpha(1-\cos\delta)&\sin^22\alpha+\cos^22\alpha\cos\delta&\cos2\alpha\sin\delta\\
0&\sin2\alpha\sin\delta&-\cos2\alpha\sin\delta&\cos\delta\\
\end{array}\right)\\
\left(\begin{array}{c c c c}
1&\cos(2\varphi)&\sin(2\varphi)&0\\
\cos(2\varphi)&\cos^2(2\varphi)&\cos(2\varphi)\sin(2\varphi)&0\\
\sin(2\varphi)&\cos(2\varphi)\sin(2\varphi)&\sin^2(2\varphi)&0\\
0&0&0&0\\
\end{array}\right) S_{in}
\end{array}
\end{equation}
where $S_{in}$ is the input Stokes vector, $S(\varphi,\alpha)$ is the generated 
Stokes vector, $\varphi$ and $\alpha$ are the 
offset angle of the LP and the retarder
respectively with respect to the positive Stokes Q axis and
$\delta$ is the retardance value of the QWP.

For calibration purposes, the LP was positioned at 0\degr, 45\degr,
90\degr and 135\degr, coinciding with the Q, U, $-$Q and $-$U
axes respectively. This ensures that all Stokes components are present at their
maximum amplitudes and with opposite sign, in order to reduce
sensitivity to offsets and other sources of contamination in the
calibration data. Although a condition number analysis of the problem
by \cite{JakobThesis2005} suggests that angle steps of 30\degr\ over a
range of only 180\degr\ should be sufficiently accurate to fully determine the
problem, the angles for the retarder were taken to be every 5\degr\
from 0\degr\ to 365\degr, a total of 74 angles, in order to average
out any error in the QWP angles and photon noise in the data that may
exist.

Strictly speaking, the modulation matrix elements can now be fitted by
a simple matrix inversion. However, this would assume the retardance
value and offset angle of the QWP are perfectly known, as well as that
the intensity was constant during the recording of the calibration
data. As the expressions for the retarder parameters are non-linear and, 
in order to compensate for the varying intensity, the observations
must be normalized to I, for which the inverse of the modulation
matrix itself needs to be known, to deal with either one of these problems,
an iterative approach is the most obvious way forward.

Using a downhill gradient-search algorithm for all parameters is the simplest
way to find the solution, but it is rather slow and can benefit
greatly from step direction optimization methods such as 
conjugate-gradient optimization. Unfortunately, it is easy to end up in a local
minimum near the global minimum.  An improvement can be made by using
as much of the explicit solution as possible ($M$) and using a downhill
search algorithm for the remaining parameters ($\alpha$ and $\delta_{QWP}$). 
Although this method
converges a lot faster and is less likely to find a local minimum, it
is more likely to end up with a completely wrong solution if the start
solution is far away from the real solution. However, both methods are
based on minimizing the same error function and converge toward the
same solution, if sufficiently restrictive convergence criteria are
used.

\begin{figure}[!ht]
\includegraphics[width=\columnwidth, bb=22 8 491 270, clip]{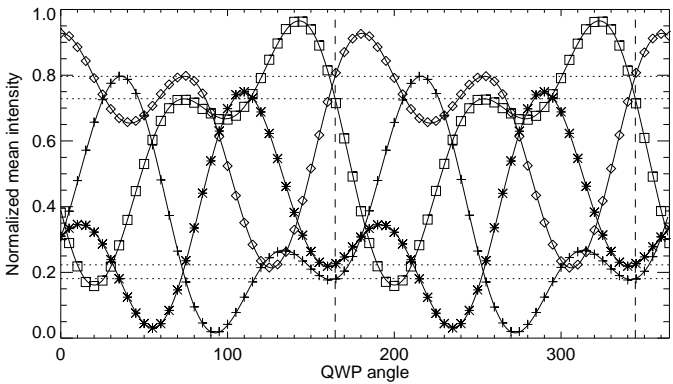}
\caption{Polarimetry Calibration for 13-Sep-2007. The lines are
  theoretical curves fitted to the data recorded for the polarimetric
  states. The horizontal dotted lines are the LP-only observations,
  the vertical dashed lines are the fitted angle offset (for 0\degr\
  and 180\degr) of the QWP. The accuracy with which the intersection
  points of the LP-only lines and their corresponding curves line up
  vertically and how well this matches the fitted offset angle is an
  indication for the overall accuracy of the fit.}
\label{fig:polcalibfig}
\end{figure}

In addition to the calibration data, observations were made without 
the QWP. This provides for an independent way of checking the quality
of the fit, as the fitted data curves should assume the value of the
LP-only observations at the retarder angle equal to that of the LP.

Figure~\ref{fig:polcalibfig} shows an example of a calibration data
set with fitted model curves, for the mean over the field of
view. Typically, a total of 8 of these curves are used to fit the
modulation matrix elements. The QWP angle for which the curves assume
the value of the LP-only observations is the same for all curves to
within $0.1^\circ$, but differs from the calibrated QWP offset angle
by $0.45^\circ$, possibly caused by inaccuracies introduced by the
manual positioning of the LP, the positions of which were not included 
in the fit.

From the modulation matrix resulting from the calibration,
efficiencies of 0.56,0.53 and 0.51 for Stokes $Q$, $U$ and $V$
respectively can be derived. The total efficiency of $0.94$ indicates
that the modulation scheme is close to the theoretical optimum.

\subsection{Telescope model}

Since the telescope uses 2 mirrors at 45\degr\ angle of incidence and
a vacuum-window, it is rather strongly polarizing and the polarization
depends on the pointing direction. In order to predict the Mueller
matrix of this system for arbitrary azimuth and elevation angles, it
needs to be modelled and then calibrated. The model is based on the
method used by \cite{1997ApJS..110..357S}, a method that has been
successfully applied to model the DST, the SVST
\citep[LPSP,][]{1997ASPC..118..366S, 1999ASPC..183..264M} and the
German VTT \citep{2005A&A...443.1047B}.  \citet{JakobThesis2005}
adapted the method to describe the SST optics.

The Mueller matrix of the telescope is the product of the
series of Mueller matrices of each of the optical components the
light passes through before exiting the bottom of the telescope
\begin{equation}
M_{tel}= M_{Schup}\cdot M_{field}\cdot R_{az}\cdot M_{az}\cdot R_{el}\cdot M_{el}\cdot L,
\end{equation}
with the main objective lens matrix $L$, the folding mirrors $M_{el}$
and $ M_{az}$, the field mirror $M_{field}$ at the bottom of the
telescope and the Schupmann corrector mirror $M_{Schup}$ respectively,
with appropriate rotations $R$ to the natural frame of the next
component.

The resulting matrix is obviously position dependent, but this is
through the rotations only, so that this dependence is known once the
matrices for all other components are known. The mirror matrices can
be described using a general expression with only 2 unknown
parameters, so that only $L$ and a a few parameters need to be
calibrated. This can only be done by fitting the predicted telescope
response to known input Stokes data.

\begin{figure}[!ht]
\includegraphics[width=\columnwidth, bb=30 8 491 348, clip]{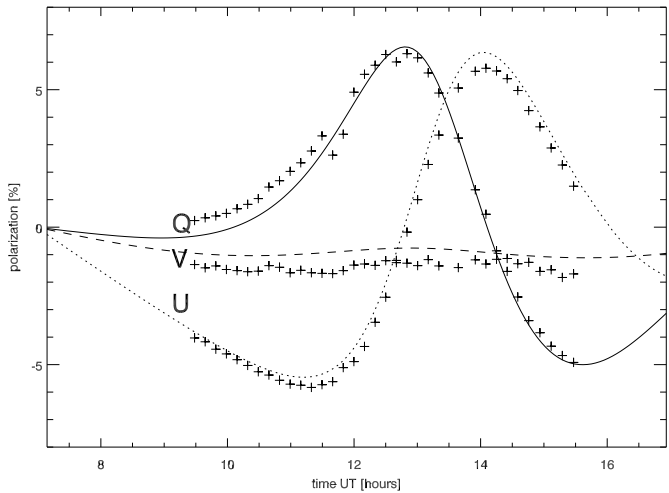}
\caption{Instrumental polarization for 12-Sep-2006 as a function of time.
The plotted curves are the model predicted telescope polarization for Q(solid),
U(dotted) and V(dashed), the crosses are the actual values measured below the 
exit window of the telescope.}
\label{fig:instpol}
\end{figure}

Unfortunately, since the entrance window has a 1\,m diameter, it is
not possible to use the same method of generating Stokes vectors as
for calibrating the polarimeter, as a 1\,m retarder of sufficient
quality is hard to come by. Instead, a fit to linearly polarized light
only will have to suffice.

To provide a source of linearly polarized light, a large sheet of
polarizing material was mounted on a rotating wheel in front of the
telescope main lens. This wheel was then slowly rotated continuously
while pointing toward the Sun, thus providing all Stokes parameters
except Stokes V. Polarimetric data was then taken continuously
throughout the day, to provide information on the azimuth and
elevation dependence of the telescope matrix.

Although of fundamental importance to the accuracy of the Stokes data
presented here, this procedure is beyond the scope of this paper and
is outlined in detail by \cite{JakobThesis2005}.  We limit ourselves to stating
that the estimated error in the predicted output of the telescope was
found to be below 0.5\%\ in Stokes Q, U and V, based on the only
calibration performed to date, made in March 2004. However, as no
information is available as to the effect of aging and maintenance on
the model parameters, this estimate may no longer be accurate.

Figure~\ref{fig:instpol} shows the instrumental polarization estimated
from a full day of Full Stokes data recorded on the 12-September-2006,
more than a year after the telescope model was calibrated. As the FOV
is dominated by a large sunspot close to disc center, the derived
offset for each Stokes parameter may contain a non-instrumental
component of solar origin.  Although the model predictions show a
small systematic difference from the data, in particular the extreme
values of Q and U appear to be assumed at a somewhat different time of
day from the model, the agreement with the telescope model appears to be 
around 1\%\ of I or better, not dramatically worse than what was found 
18 months earlier.

\section{Observations and data reduction}
\label{sec:obs}

\begin{figure*}[!ht]
\includegraphics[width=\textwidth, bb=5 0 823 315, clip]{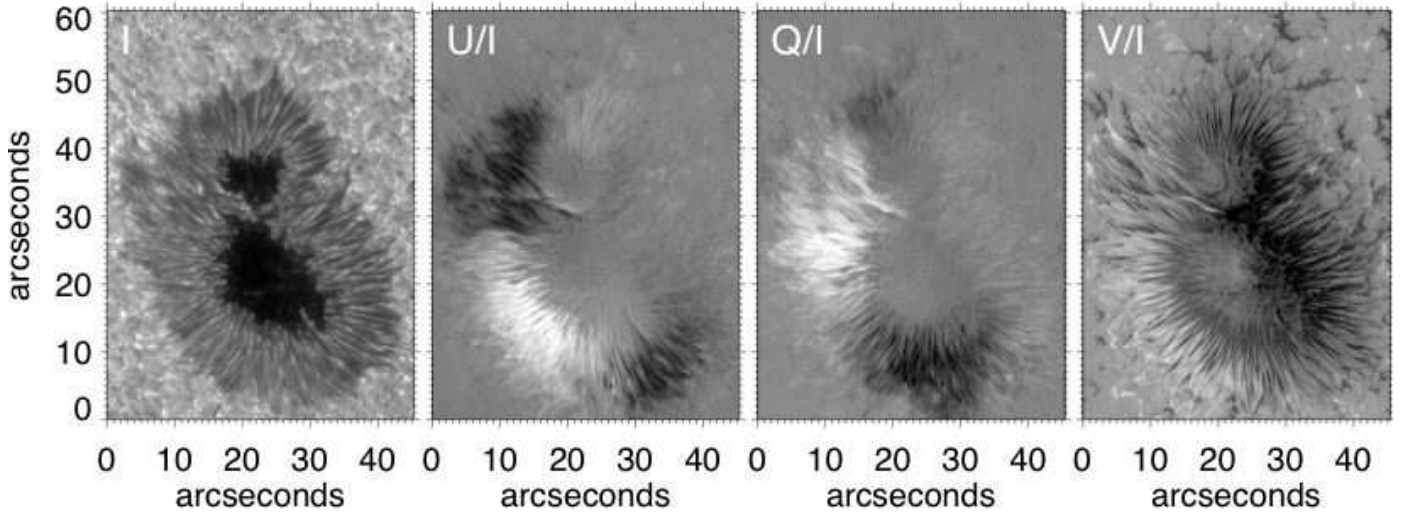}
\caption{Sunspot AR10908 observed on 12-Sep-2006 in
  \ion{Fe}{I}~630.25~nm $-$5~pm. The color scaling saturates at
  $\pm15$\%\ for U/I and Q/I, and at $\pm25$\%\ for V/I.}
\label{fig:overview}
\end{figure*}

From several campaigns during the 2006 observing season, we selected 4
data sets of sunspots that were taken under excellent seeing
conditions. 
Detailed information on the different targets is presented in
Table~\ref{table:obs}. 
Figure~\ref{fig:overview} shows the four Stokes maps for
\ion{Fe}{I}~630.25~nm $-$5~pm for one of the selected sunspots.

The quality of the data benefited from the SST adaptive optics system
\citep[AO,][]{scharmer2003AO} and post-processing using the
Multi-Object Multi-Frame Blind Deconvolution
\citep[MOMFBD,][]{vanNoort05MOMFBD} image restoration method.
The use of MOMFBD is complementary to the use of AO in a sense that it
provides the possibility to achieve high image quality also at large
angular distance from the AO wavefront sensor. 

\begin{figure}[!htb]
\includegraphics[width=\columnwidth]{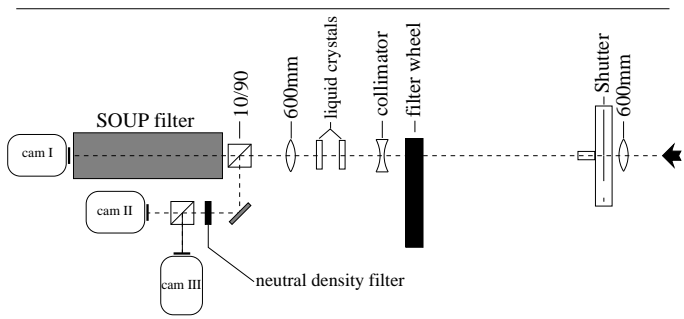}
\caption{Schematic drawing of the optical set-up.}
\label{fig:optsetup}
\end{figure}

Narrow band images in the wings of the \ion{Fe}{I}~630.25~nm line were
obtained with the Solar Optical Universal Polarimeter
\citep[SOUP,][]{title81SOUP} which has a FWHM of 7.2~pm.
Three science cameras were employed with 1024$\times$1024 pixel CCD
chips (pixel scale 0\farcs063, the diffraction limit $\lambda/$D of
the SST at 630 nm is 0\farcs13).
One of these cameras was positioned behind the SOUP filter, the other
two were set-up as a phase-diversity pair that received 7.5\% of the
light on a branch that split off the main beam before the SOUP
but after the SOUP pre-filter (0.8~nm passband centered on
630.25~nm). 
The exposures of all three cameras were synchronized by means of an
optical chopper.
The optical setup was similar to the one described in
\citet{depontieu2007DFs}, with an additional pair of LCVRs positioned
between the collimator and re-imaging lens, and is illustrated in
Fig.~\ref{fig:optsetup}.

As the polarimetric signals of interest are usually weak, it is of
crucial importance to obtain the largest S/N ratio possible. Although
the Sun is an excellent source of photons, when reducing the spectral
range to only a few tens of pm, acquiring a detectable quantity of 
photo-electrons in the few ms permitted by the seeing 
becomes a challenge. 
In order to make the most of the available light, the cameras used at
the SST are Sarnoff 1M100 cameras, the CCD of which has a quantum
efficiency of 0.7 at 630~nm and which consists of 16 separate segments
that can be read out in parallel. By using a slow pixel clock, this
provides for a camera with exceptionally low readout noise as well as
a readout time of only 10~ms.  Even with this short readout time, the
most effective exposure time is a matter of compromise, as from the
seeing point of view it should be as short as possible, but from the
S/N standpoint it should be at least somewhat longer that the readout
time in order to get a reasonable efficiency. In practice the rotation
rate was set to $\sim$36~Hz, giving an exposure time of $\sim$17~ms
and a duty-cycle of approximately 60\%.
Apart from the desire to have a large duty-cycle, a shorter exposure
at the current light levels (using the SOUP filter which has a
transmission of only $\sim$15\%) would result in a photon noise that
is at or below the readout noise of the camera, a situation that the
MOMFBD reduction is currently unable to handle.

The LCVRs of the polarimeter were cycling through a sequence of 4 states, 
changing state for each exposure during the read-out time of the cameras. 
The SOUP filter was running a program switching between the red and
blue wing of the \ion{Fe}{I}~630.25~nm line, at $\pm5$~pm offset from
the line core.
The change of line position was done after about 14~s, covering typically
125 LCVR cycles. 
With a line position changing time of about 7~s, the acquisition of a
full data set containing two line positions and all LCVR cycles took
about 35~s.
The exact number of images (and effective exposure time) for each data
set is specified in Table~\ref{table:obs}. 

\begin{figure}[!ht]
\includegraphics[width=\columnwidth, bb=0 0 570 500, clip]{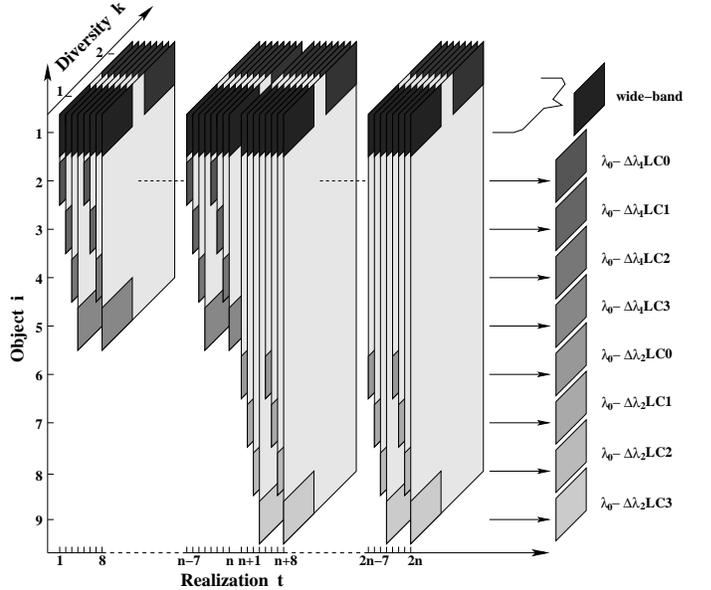}
\caption{Sketch of the acquisition scheme of a full Stokes MOMFBD data
  set for 2 line positions. The 2 line positions,
  $\lambda_0-\Delta\lambda_1$ and $\lambda_0-\Delta\lambda_2$, are
  recorded sequentially. For each line position, the LCVRs are cycling
  through 4 states LC0--LC4, changing state for each exposure t. For
  each narrow-band exposure, a pair of accompanying wide-band
  phase-diversity images are recorded. The wide-band channel serves as
  anchor channel for the narrow-band images. For the data sets
  discussed in this paper, the number of exposures for each line
  position is 500, i.e., n=500. }
\label{fig:momfbd-sketch}
\end{figure}

Figure~\ref{fig:momfbd-sketch} illustrates the acquisition scheme of
the data. Since the cameras are synchronized, 3 exposures share the
same wavefront realization $t$ (x-axis): one SOUP image with a
specific line position and LCVR state $o$ (z-axis) and two wide-band images with
known diversity $k$ (y-axis), one in focus and one slightly out of focus 
($\sim$12~mm on a 45~m focal length).

The figure is similar to Fig.~7 in
\citet{vanNoort05MOMFBD} with the important difference that the LCVR
states are now modulated for each exposure instead of after a burst
of several exposures.
This results in a collection of polarimetric states that is more
homogeneous in terms of seeing aberrations and means that the level of
seeing cross-talk is highly reduced already before MOMFBD processing.
The interleaved character of the data set is akin to the high
speed modulation used by the Z{\"u}rich Imaging Polarimeter (ZIMPOL,
\cite{1997A&A...328..381G}), which employs modulation in the kHz range, 
and is highly beneficial for the construction of high-sensitivity Stokes 
maps.

While the images from the SOUP camera are constantly varying due to the
modulation of the LCVRs and the changing line position, the images from
the wide-band cameras remain unchanged other than by seeing variations. 
Since the wavelength of the wide-band images is nearly identical to that of
the narrowband images, the wide-band images can be used as an "anchor" channel, 
to which all the restored SOUP objects are aligned in the MOMFBD restoration 
procedure, as discussed at length in \citet{vanNoort05MOMFBD}.

To calibrate the inter-camera alignment, images were recorded of an array of 
30~$\mu$m pinholes. By fitting these pinhole images to each other under the 
assumption that they look the same, the relative offset of each camera pixel 
with respect to a properly chosen reference camera can be calculated using the 
MOMFBD code in calibration mode. These image offset calibrations result in 
alignment of restored objects with a typical accuracy of $\approx$0.05 pixels.
This means that the restored SOUP objects have
near-perfect alignment -- even though the individual exposures are not
recorded simultaneously -- and no further alignment by means of
cross-correlation is needed to construct the Stokes maps. 
The level of false signals induced by seeing is therefore generally
very low (see also Sect.~\ref{sec:noise}). 

\begin{table}
  \caption{Summary of observed targets. $\mu=\cos(\theta)$ with
    $\theta$ the observing angle. e time is the effective exposure
    time per line position, i.e., the sum of all exposure times.}
  \label{table:obs}
  \centering
  \begin{tabular}{ccccc}
    \hline\hline
    date & NOAA number & position [$\degr$] & $\mu$ & e time \\
    \hline
    07-Jan-2006 & AR10845 & N18\,E64 & 0.39 & 7.8\\
    10-Jun-2006 & AR10893 & N01\,E17 & 0.95 & 7.8\\
    12-Sep-2006 & AR10908 & S09\,W12 & 0.94 & 7.5\\
    15-Sep-2006 & AR10908 & S13\,W51 & 0.58 & 7.7\\
    \hline
  \end{tabular}
\end{table}

Unless otherwise stated, all images were reduced using 128x128
(5x5\arcsec) and seamless mosaicking of the tapered subimages. The
reductions were done on a 90 CPU 3.2 GHz Intel Xeon cluster at the
University of Oslo using 34 Karhunen-Lo\'eve wavefront modes and
converged until the relative change in the wavefront coefficients was
$<10^{-4}$.

\section{Quality Assessment}

\subsection{Noise}
\label{sec:noise}

One of the main motivations for conducting the observations the way we
describe in this paper is the detection of weak polarization
signals. It is therefore important to get a firm handle on the noise
level in the data. The main source of noise is photon noise. Read-out
noise is a minor source of noise: for the Sarnoff cameras the read-out
noise is specified by the manufacturer to be on the order of 25 electrons, 
which amounts to 1--2 counts. Seeing cross-talk is another potential source
of noise. 
As discussed in Sect.~\ref{sec:obs}, the use of a wide-band anchor
channel in MOMFBD restorations greatly minimizes seeing cross-talk in
derived quantities like magnetograms. In addition, the use of
interleaved LCVR state exposures reduces the effect of seeing
cross-talk even more.

Even though, in order to investigate the noise, images free from polarization
signal are needed, there is no location on the Sun that is guaranteed
to be completely free of magnetic field. Moreover, observing at a
presumably field-free location or at a magnetically insensitive
wavelength does not provide polarization free images anyway due to the
instrumental polarization cross-talk. Although the latter can be
corrected for using the telescope model, this potentially entangles
the polarization cross-talk with the seeing cross-talk, which is not
desirable.

Instead, we opted for "re-labeling" the data recorded for one specific
modulation state and wavelength to the other states and reducing this
dataset as if it were polarized data. Since the signal in all images
corresponds to the same state, no polarization signal should be
detected after the reductions are complete. By repeating this for all
states, a reasonable estimate can be made of the false signal due to
noise, seeing-crosstalk, etc. Unfortunately, since the polarimeter
itself is also polarizing in Stokes Q and U, but not much in V, this
method could only be used on data that contain only Stokes V
observations.

\begin{figure}[!ht]
\includegraphics[width=\columnwidth]{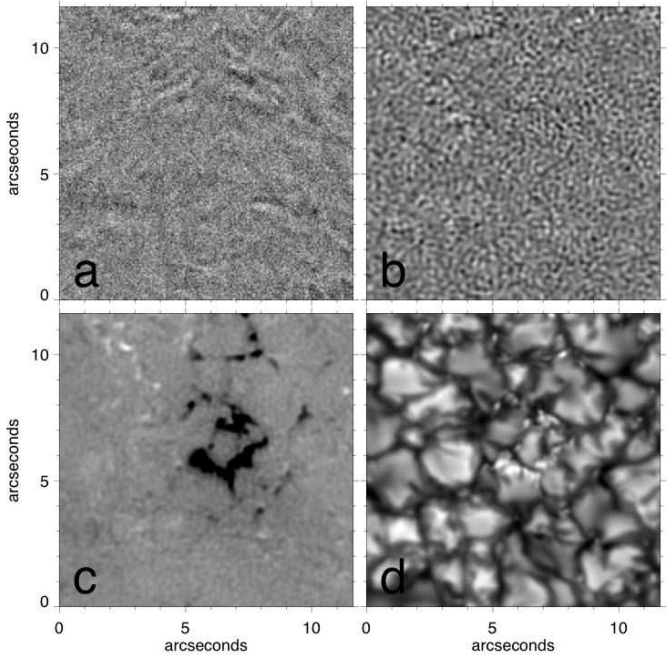}
\caption{Panel (a) shows a magnetogram-like image constructed from
  flatfielded images of the same liquid crystal state. Panel (b) shows
  a similar image constructed from a MOMFBD restoration based on the
  same images as panel (a). Both panel (a) and (b) have the same
  scaling, $-0.01$ to $+0.01$. Panel (c) shows the \ion{Fe}{I}~630.2
  $-$5~pm stokes V magnetogram from the same data set and the same
  number of images. The scaling is $-0.05$ to $+0.05$. Panel (d) shows
  the wide band image of the same region and data set.}
\label{fig:noiseimages}
\end{figure}

The top panels of Fig.~\ref{fig:noiseimages} show parts of such
``fake'' magnetograms constructed using the standard formula
$(I_{LCP}-I_{RCP})/(I_{LCP}+I_{RCP})$ but with $I_{LCP}$ and $I_{RCP}$
being from the same state. Panel (a) is constructed from flat-fielded
images, panel (b) is based on MOMFBD restored data. Since the FOV
shown in these panels is off-center, away from the AO wavefront sensor
target area, anisoplanatism can be expected to introduce image shifts
and deformations that introduce seeing crosstalk. Indeed, for the
flat-fielded data we see patterns on granular scales that can be
attributed to seeing crosstalk. For the MOMFBD data, no such patterns
can be discerned. This is confirmed by a power spectrum analysis which
shows no conspicuous bumps that indicate a significant level of false
signal due to seeing crosstalk.

The noise pattern in panel (a) and (b) is rather different: the
``orange-peel'' pattern for the MOMFBD data has a larger
characteristic scale than for the flat-fielded data which is varying
on a pixel-to-pixel scale. The characteristic ``orange-peel'' pattern
in the MOMFBD data can be explained by the low-pass filter employed
during the restoration:
power at the highest frequencies, beyond the diffraction limit and the
noise-dominated regime in Fourier space, are attenuated. This leaves a
background or noise pattern with a characteristic spatial scale of a
few pixels.
The noise filter is a fundamental part of the deconvolution conducted in
the MOMFBD procedure. It actually suppresses the noise level slightly
below the level one expects from photon noise (see below). However,
the characteristic noise pattern with a scale close to the diffraction
limit poses problems for the detection of small-scale and weak-signal
features as it is difficult to make an unambiguous distinction from
noise at that scale and low signal level. In such cases, a time
sequence of measurements could validate a detection of weak features,
provided that they survive sufficiently long.
 
Uncertainties in the knowledge of the varying photo-sensitivity of the
detector pixels is another potential source of error. This kind of
noise is also known as gain table or flat-field noise. The same
detector is used for recording all polarization states but the
flat-field for each state is found to be different and therefore
uncertainty in the determination of the flat-field introduces noise.
For our observations, flat-fields are constructed by adding a large
number of exposures while the telescope is sweeping over the solar
disk.
Due to the rapid motion and addition of a large number of images,
solar features are effectively averaged out. We find that by adding
more than 1000 images, the uncertainty in the flat-field is below
10$^{-3}$ such that this kind of noise is negligible.

\begin{figure}[!ht]
\includegraphics[width=\columnwidth]{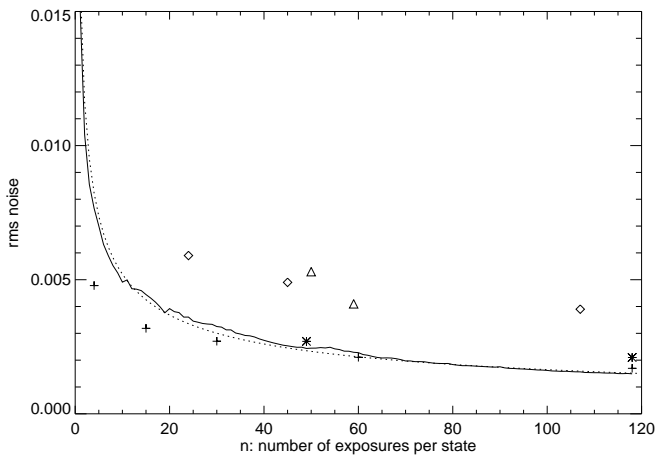}
\caption{Noise in ``fake''magnetograms as function of number of
  exposures per state. The solid line is for flatfielded data, similar
  to panel (a) of Fig.~\ref{fig:noiseimages}. Crosses are MOMFBD
  restored data, similar to panel (b) of Fig.~\ref{fig:noiseimages}. The
  dotted line is the 1/$\sqrt{n}$ noise behaviour of the photon-noise
  dominated case with 17 photo-electrons per count. Noise measurements
  on real -50~m\AA\ magnetograms: quiet Sun Stokes V only (asterisks),
  Stokes Q close to the sunspot of 12-Sep-2006 (diamonds), Stokes U
  disk center from 2007 (triangles). }
\label{fig:noiseplot}
\end{figure}

Figure~\ref{fig:noiseplot} shows the RMS noise level of several
restored data sets as a function of the number of frames per state,
for images "restored" using flat-fielding and addition only and for
images restored using MOMFBD reduction. The flat-fielded data follows
the expected 1/$\sqrt{n}$ behaviour very well which confirms that the
exposures are dominated by photon noise. The noise level of the MOMFBD
restored images is actually lower than the theoretical 1/$\sqrt{n}$
limit, presumably the result of reduced high-frequency noise from
filtering beyond the diffraction limit.

A number of noise measurements of real Stokes magnetograms are plotted
in Fig.~\ref{fig:noiseplot}. These measurements were done on small,
presumably field-free regions in the FOV. The noise in full Stokes
measurements is about two times higher than in observations devised to
measure only Stokes V. This is because the efficiency to measure the 4
Stokes parameters is four times lower as compared to measuring Stokes
V only.

For a full Stokes measurement, 125 exposures per state, or
500 exposures in total, is about the maximum number of exposures that
can be included if one considers the photospheric evolution time. It
takes about 14~s to acquire these 500 exposures. A disturbance moving
at a velocity of the photospheric sound speed ($\sim$7~\kms) travels
about 100~km during that time, which is about the distance of the
diffraction limit at this wavelength.

Following the same arguments for a measurement of only Stokes V,
double the amount of exposures per state can be used since the liquid
crystals alternate between two states only. In the Stokes V only data
set used for Fig.~\ref{fig:noiseplot}, we measure a noise level of
about $2\cdot 10^{-3}$ in \ion{Fe}{I}~630.2~$-5$~pm. Using the
calibration constant of 16551~Mx~cm$^{-2}$ per polarization percent,
found by \citet{2004A&A...428..613B} for \ion{Fe}{I}~630.2~$-$5~pm SOUP
stokes V magnetograms, this noise level is equivalent to a magnetic
flux density of 33~Mx~cm$^{-2}$.

\subsection{Spatial resolution}

In order to establish the achieved spatial resolution of the
observations, we do a power spectrum analysis and a close inspection
of the smallest resolved structures in the data. 

\begin{figure*}[!ht]
\includegraphics[bb= 0 0 500 162, clip, width=\textwidth]{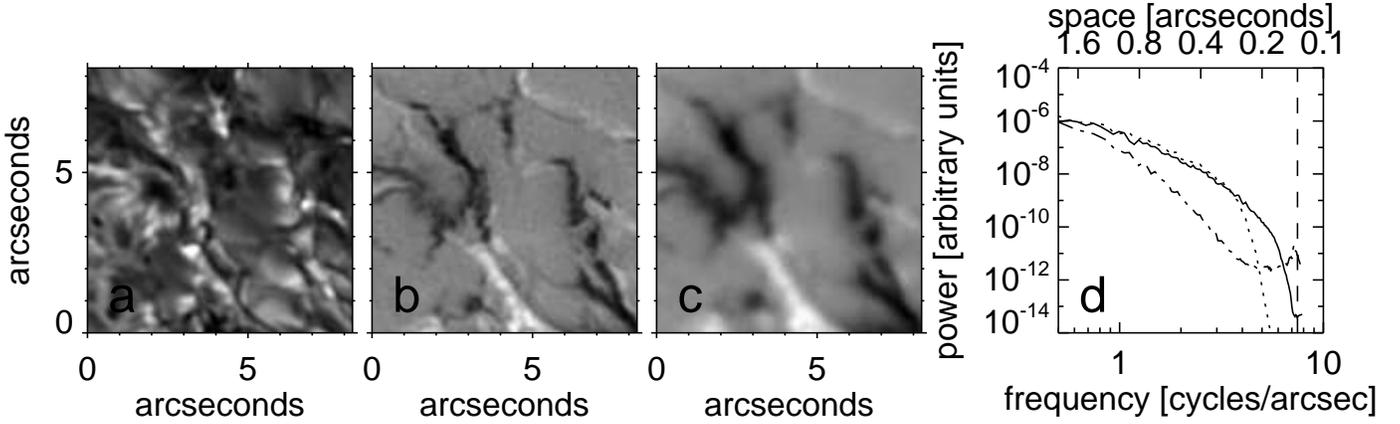}
\caption{Panel (a) shows a part of a \ion{Fe}{I}~630.2~$-5$~pm Stokes
  I image, observed on 12-Sep-2006, MOMFBD restored based on 110
  exposures per state. Panel (b) shows the corresponding MOMFBD
  restored stokes V image. Panel (c) is a Stokes V image constructed
  from the flat-fielded data. Panel (b) and (c) have the same scaling,
  -0.2 to 0.2. The graph in panel (d) shows power spectra of part of
  the MOMFBD restored wide-band image (solid line), the MOMFBD
  restored Stokes I image (dotted), and the flat-fielded stokes I
  image (dash-dotted line). The vertical long-dashed line marks the
  theoretical diffraction limit $\lambda$/D of 0\farcs134 at this
  wavelength.}
\label{fig:resolution}
\end{figure*}

Figure~\ref{fig:resolution} shows sample images of the observations
from 12-Sep-2006. 
The images shown are based on 110 exposures per state, meaning a total
of 440 exposures. The data for panel (c) has been simply flatfielded
and added, as if it were obtained using long exposures (1.65~s per
state, 6.6~s effective exposure in total), but still with the added
bonus that the states are interleaved and therefore the seeing for
each of the states is more equal than for real long exposures.
The difference in resolution between panel (b) and (c) is striking. The
MOMFBD restored Stokes V image shows a lot more fine detail and there
are many examples visible of small, weak features that are not
resolved in the flat-fielded stokes V image. 

The power spectra in panel (d) are angular averages of 2D Fourier
power spectra of a 256$\times$256 pixel sub-area in a corner of the
FOV, away from the sunspot and showing only granulation.
The power spectrum of the MOMFBD restored wide-band image (based on
440 focus and 440 de-focus images) shows a gradual decrease in power
to higher frequencies, with a steeper decline between about 0\farcs18
and just above the diffraction limit $\lambda$/D at 0\farcs134. The
steep decline is an effect of the noise filter that has a smooth
transition from no effect on the lower frequencies towards full
suppression of the highest, noise dominated, frequencies.
The SOUP exposures that are used to construct the Stokes data are
noisier than the wide-band images and therefore the noise filter sets
in at a lower frequency, corresponding to a spatial scale of about
0\farcs2. 
The power spectrum for the flat-fielded Stokes I image is shown for
comparison and it illustrates how the MOMFBD restoration enhances
power at all relevant spatial frequencies. 
Note the small enhancement in power just above the diffraction limit
for the flat-fielded data. That is caused by a small localized power
enhancement in the Fourier domain. This is probably caused by some
internal camera property.  This has no significant effect on the
restored data since it is effectively suppressed by the noise filter.

It is difficult to extract a hard number for the achieved spatial
resolution from the averaged power spectrum. Even though the steep
decline due to the noise filter starts slightly above 0\farcs2, a
claim of a spatial resolution of better than 0\farcs2 in the Stokes
data is justified. We have performed a close inspection of the
smallest features in the Stokes magnetograms and we find many examples
of structures that have widths with a FWHM of about 3 pixels
(0\farcs19).

\subsection{Unresolved issues}
Even though the most obvious sources of error have been discussed
above, additional errors from limitations in instrumentation and data
processing remain. Although the observing scheme was set to minimize
these effects wherever possible, several issues remain unresolved and
should be considered when interpreting the data in detail.

\subsubsection{Doppler crosstalk}

Since, for reasons of cadence and detection sensitivity, the data was
recorded only at 2 wavelengths, a potentially serious source of false signal 
is introduced by Doppler shifting of the line profile due to motions in the 
solar atmosphere. Due to the lack of wavelength information, correcting for 
this involves assumptions regarding the actual line profile. 

Although it is possible to remove the polarized part of the line profile
using the Stokes inversions, it remains important to remember that
the line used here is magnetically sensitive and even the non-polarized 
profile may have a complicated shape. Even the only plausible assumption,
that of a symmetric line profile, may therefore only be of limited value.

Unfortunately, fake magnetogram tests, like those done in section
\ref{sec:noise}, cannot be used to get an indication of the error made
in this way, as the velocity structure evolves on the same time scale
as the solar structure, and so the line is similarly affected in both
restored images in the set.

A comparative study of the Doppler velocities obtained using a
magnetically insensitive spectral line and a magnetically sensitive
one, such as \ion{Fe}{I}~630.2, appears to be the best way forward in
resolving this issue, but will certainly require more than two
wavelength points.

\subsubsection{Restoration effects}

Application of the MOMFBD restoration method enables the removal of a
significant amount of seeing-induced aberrations from the observations
as well as sub-pixel alignment of the different objects. However, the
use of this particular image restoration technique introduces a new
kind of variability to the problem: that caused by the quality of the
restoration. This quality is to some extend still related to the
seeing quality, but also to the amount of detail used in the wavefront
sensing step of the image restoration process.

In particular, the number of atmospheric wavefront modes
(Karhunen-Lo{\`e}ve polynomials) used to fit the wavefront is of
importance. The higher order modes do not significantly affect the
core structure of the PSF, and therefore the morphology of the result,
but they have a significant impact on the wings of the PSF. This in
turn has a significant effect on the contrast of the image, similar to
the contribution of stray light.

\begin{figure}[!ht]
\includegraphics[bb= 5 9 354 339, clip, width=.495\columnwidth]{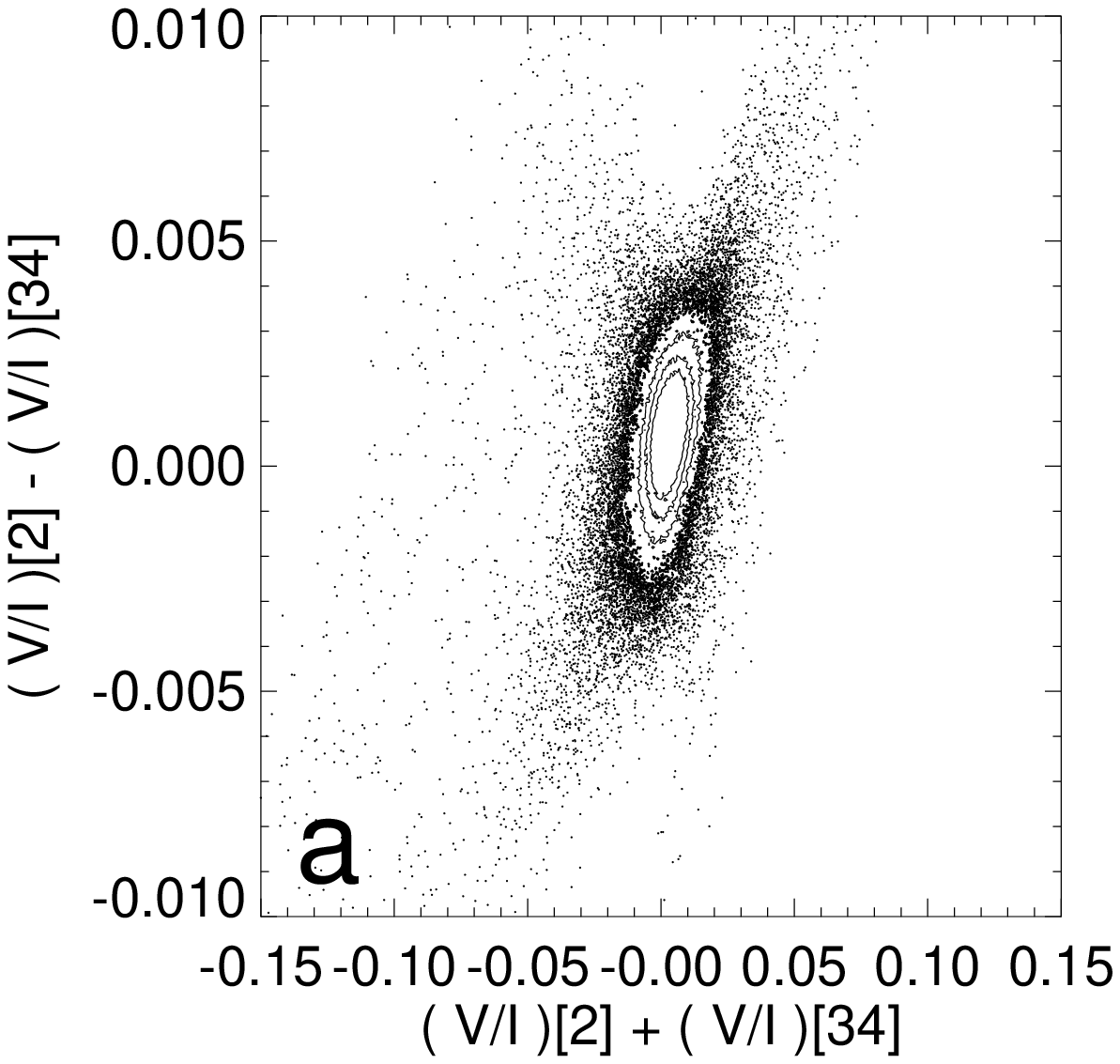}
\includegraphics[bb= 5 9 354 339, clip, width=.495\columnwidth]{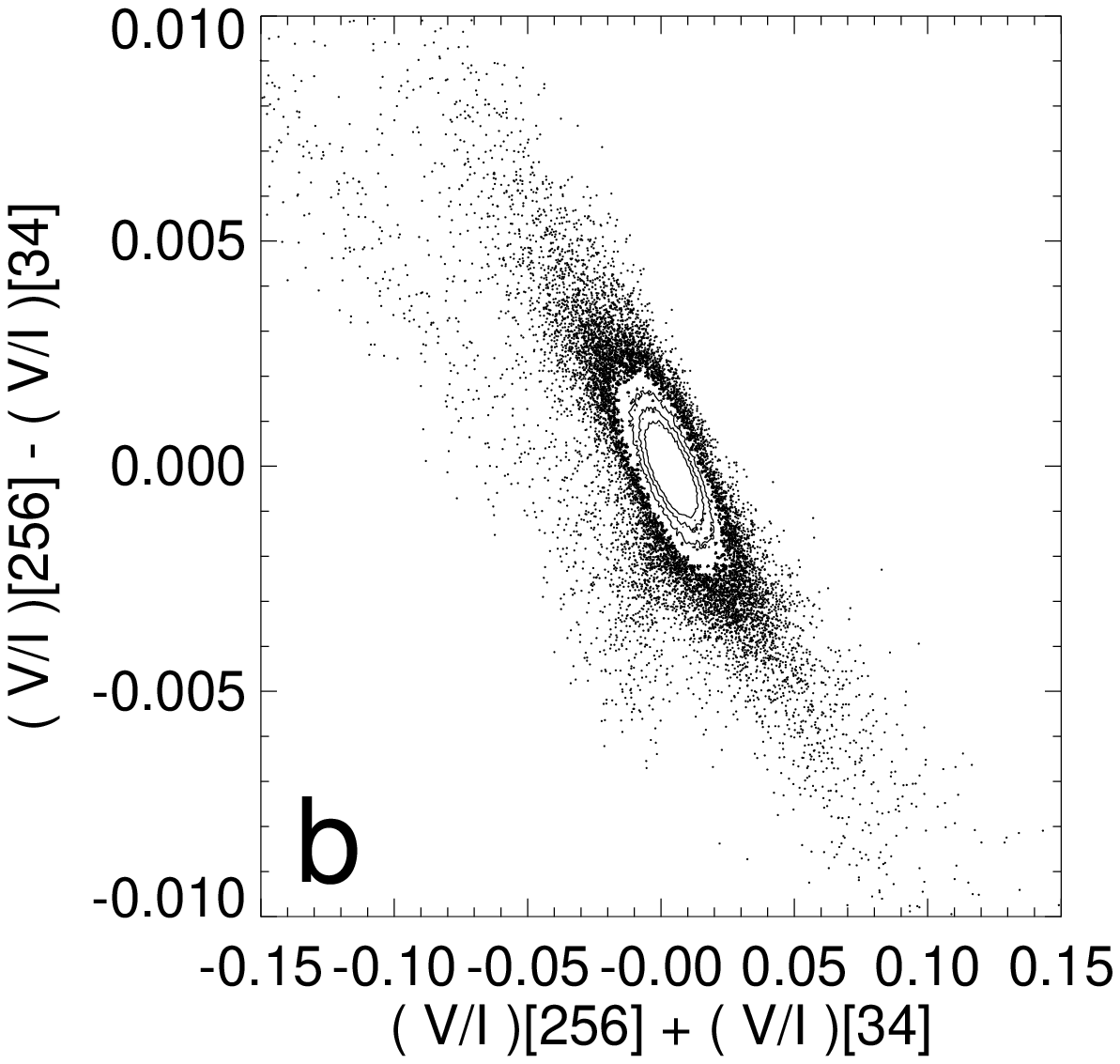}
\caption{Scatter plots of relative difference comparing Stokes V
  magnetograms restored with different number of wavefront
  modes. Panel (a) compares a 2-mode with a 34-mode restoration, (b)
  compares a 256-mode with a 34-mode restoration. For the densest
  scatter regions, density contours are plotted to avoid crowding.}
\label{fig:modes}
\end{figure}

The impact of including higher order wavefront modes on derived
quantities, like Doppler velocity and Stokes maps is therefore in
general an increase of the amplitude of the signal. This is
illustrated by Figure~\ref{fig:modes}, that shows plots of the
intensity for the same Stokes V magnetogram, restored with different
numbers of modes.

Panel (a) compares a 2-mode restoration (tip-tilt only) with a 34-mode
one which is the most common number of modes used in MOMFBD
restorations.
There is a weak correlation, with significant scatter, especially
for low-signal pixels. This is consistent with a considerable change
in the structure of the image, presumably due to an increase in the
resolution of the image caused by the removal of seeing induced
aberrations.

Panel (b) shows the same plot but comparing the 34-mode restoration
with a 256-mode one. Clearly the correlation for most pixels in the
image is tighter and can be fitted to a line with a slope of
approximately -0.1. This is more consistent with a systematic scaling
of the 34-mode image to the 256-mode image, with a scale factor of
approximately 1.2, than with a substantial change in the image
structure.

Although the effect of a scale factor on the fitted values of
atmospheric quantities resulting from detailed atmospheric modeling
may be quite limited, as the line shape should not be affected, since
no actual modeling has been performed on this data, the true error
introduced by this effect remains unknown.

\section{Results}
\label{sec:results}

\subsection{Dark-cored penumbral filaments}

Dark-cored penumbral filaments \citep{2002Natur.420..151S} are
characterized by bright edges separated by a dark central core. 
Their rather symmetrical appearance (i.e., limited fine structuring
along their length), sometimes considerable length \citep[over
6000~km, ][]{2004A&A...414..717R}, and coherent dynamical evolution and
lifetime ($\le$45~min reported by \citet{2004A&A...424.1049S} and
$\le$90~min by \citet{Langhans:2007fk}) %
all suggest that the dark-cored filaments are single entities and
constitute important building blocks of sunspot penumbrae. 
An explanation of the existence of dark cores is crucial for the
understanding of penumbrae and requires a full observational
characterization which demands for polarimetric measurements at a
spatial resolution of about 0\farcs3 and better.

For the sunspots we observed, we clearly resolve the dark cores of
penumbral filaments in the Stokes maps.  We observe that dark-cored
penumbral filaments appear vividly in polarized light, and we note
that the dark cores exhibit weaker polarization signal than their
edges.
This is readily observed in the boundary region between umbra and
penumbra, where dark-cored filaments can be unambiguously recognized.

\begin{figure*}[!ht]
\includegraphics[bb= 0 58 934 243, clip, width=\textwidth]{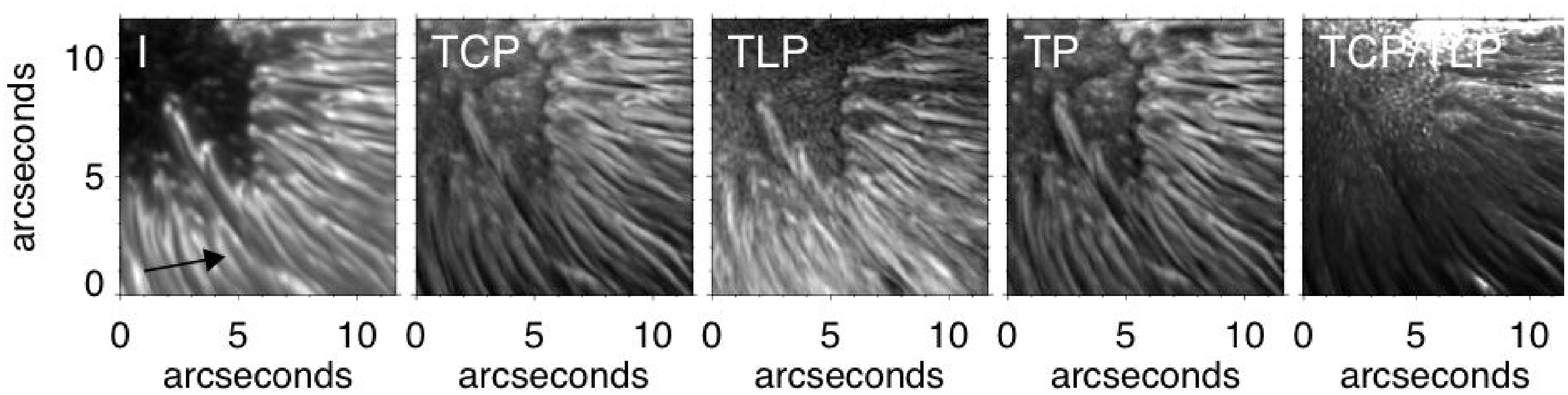}
\includegraphics[bb= 0 58 934 243, clip, width=\textwidth]{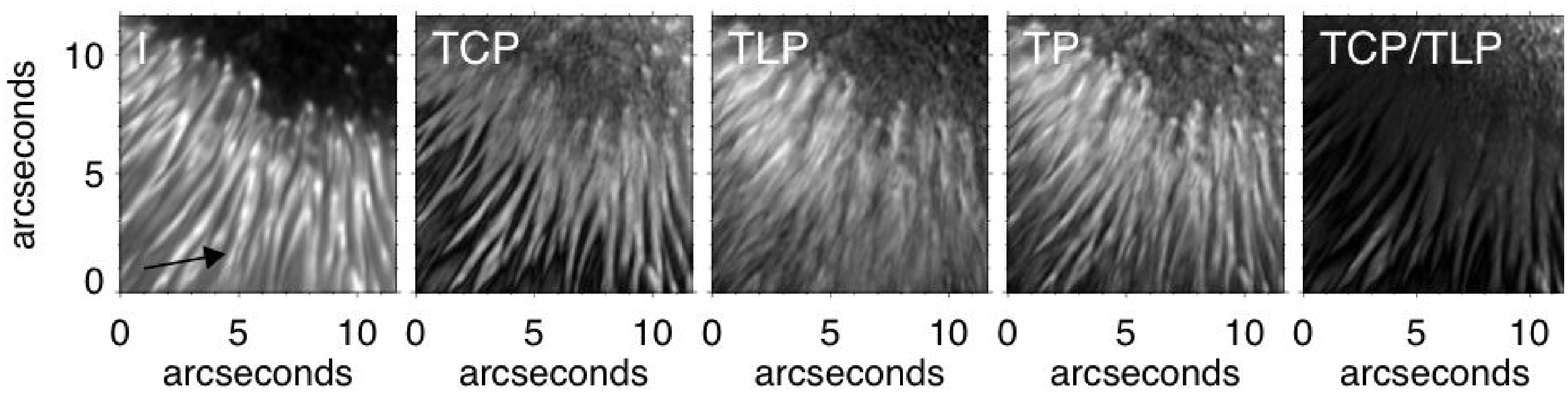}
\includegraphics[bb= 0 58 934 243, clip, width=\textwidth]{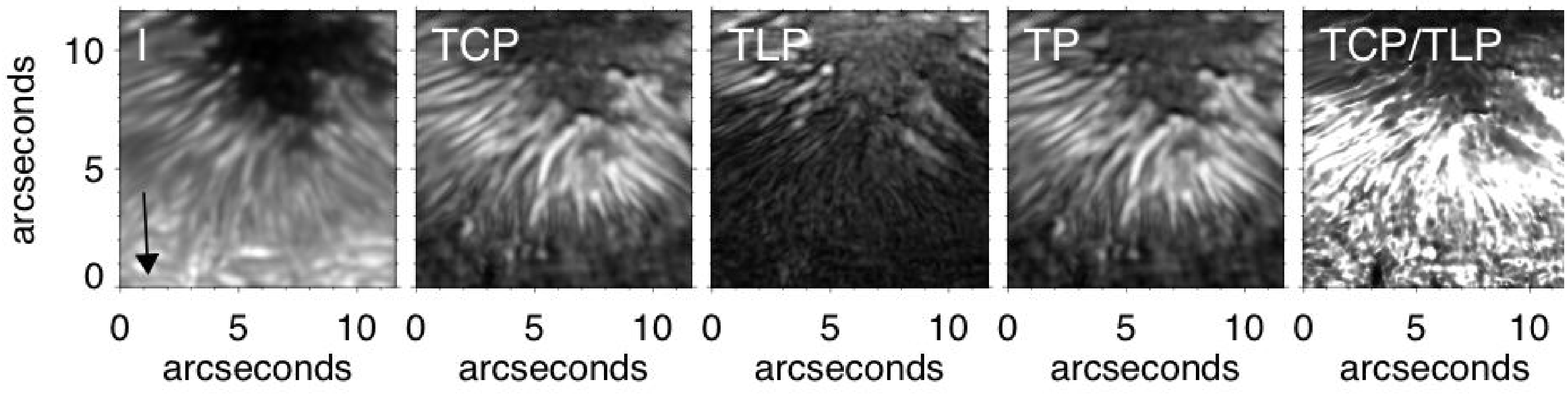}
\includegraphics[bb= 0  0 934 243, clip, width=\textwidth]{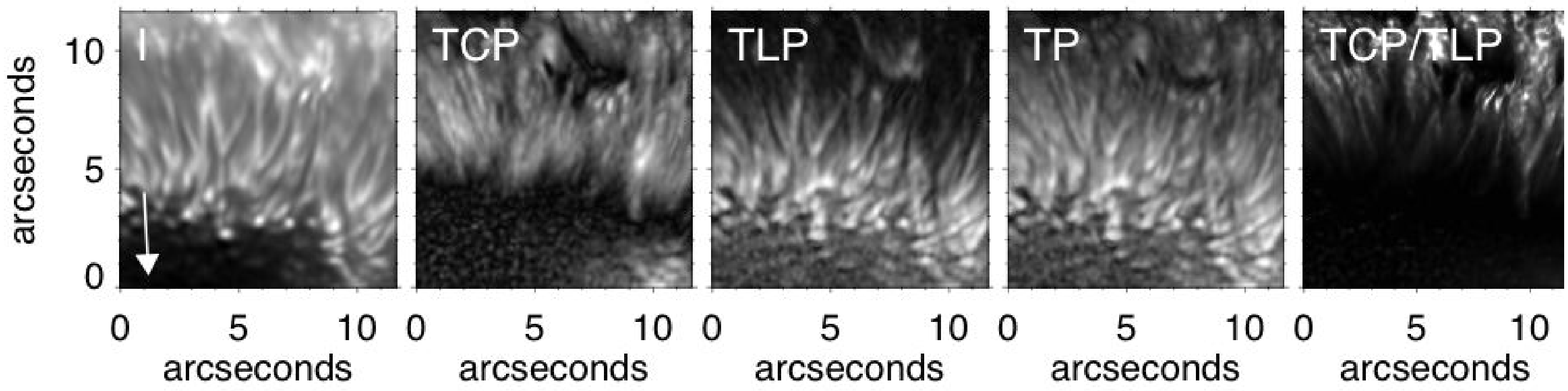}
\caption{Dark-cored penumbral filaments observed in two sunspots. Top
  two rows: AR 10908 on 12-Sep-2006, bottom rows: AR 10908 on
  15-Sep-2006, with the center-side penumbra on the third row, and the
  limb-side penumbra on the bottom row. Left to right: maps of Stokes
  I in $+$5~pm, total circular polarization, total linear
  polarization, total polarization, and the ratio of total circular
  over linear polarization. The arrow in the Stokes I images points
  towards disc center.}
\label{fig:darkcores}
\end{figure*}

Figure~\ref{fig:darkcores} shows examples from several datasets: maps
of total circular polarization TCP, total linear polarization TLP, and
total polarization TP. Traditionally, these maps are constructed by
integration over wavelength. Since we have only 2 wavelength
positions, these maps are constructed by summing:
TCP$=\sum |V_i|/I_i$, 
TLP$=\sum (Q_i^2 + U_i^2)^{1/2}/I_i$, and 
TP$=\sum (Q_i^2 + U_i^2 + V_i^2)^{1/2}/I_i$,
where the summation is performed over two wavelengths $i$:
$\pm$~5~pm.
We see that the dark cores have weaker signal in all summed maps. This
is an indication that dark cores possess weaker magnetic fields than
their edges. 

The top two rows show maps from a spot close to disk center, observed
on 12-Sep-2006. 
These maps show many clear examples of weaker signal in dark cores.
The bottom two rows are from a sunspot observed towards the limb at an
angle $\theta=55\degr$ ($\mu=0.58$). There is an interesting
difference between limb-side (bottom row) and center-side penumbra:
The center-side TLP map shows only very weak signal whereas the
limb-side TLP map distinctively shows high contrast between dark cores
and their edges.
This must be interpreted as a line-of-sight (LOS) effect with the magnetic
field having small inclination angles with respect to the LOS on the
center-side penumbra. 
This means that the angle of the magnetic field with respect to the
solar surface must be about $35\degr$ for the penumbra of this
sunspot. 
That would also explain the relatively weak signal in TCP for the
limb-side penumbra as compared to the center-side.  

In the last column of Fig.~\ref{fig:darkcores}, maps of the ratio of
circular over linear polarization are shown. 
The dark cores in the top two maps of the 12-Sep-2006 spot appear
rather clearly and they have a relatively enhanced TLP signal as
compared to their edges (i.e., the dark cores stand out as dark
features). 
This indicates that the magnetic field in dark cores have higher
inclination angles with respect to the normal as compared to their
bright edges.  For the center-side TCP/TLP map, the TLP signal is too
low to show any significant features.
In the limb-side map, the dark cores stand out as {\it bright}
features. 
This is also compatible with higher inclined fields in the dark cores
since for this part of the penumbra that implicates a lower
inclination with respect to the LOS and therefore enhanced TCP
signal. 

These observations of weaker signals in dark cores cannot be the basis
for a firm conclusion on the magnetic field strength and orientation
in dark-cored penumbral filaments. A definite conclusion must be based
on thorough modelling that takes the thermodynamic properties of the
penumbral atmosphere into account. These observations support the
findings from \citet{Langhans:2007fk} and \citet{2007ApJ...668L..91B}.
\citet{Langhans:2007fk} concluded on the basis of a geometrical study
on high spatial resolution SST Stokes V magnetograms that dark cores
possess weaker field and lower inclination than their bright edges.
We achieve similar resolution and support that finding with Stokes U
and Q observations and include a compensation for Doppler shift
effects by using two wavelength positions in both wings of the
spectral line.
\citet{2007ApJ...668L..91B} find weaker dark core polarization signals
in \emph{Hinode} spectro-polarimetric observations. Their observations
have superior wavelength coverage and slightly better signal-to-noise.
Their resolution is about a factor 2 worse as compared to our
observations which means that the penumbral dark cores are barely
resolved.  Their inversions on the \emph{Hinode} data indicate that
dark cores have weaker and more inclined magnetic fields.  From the
splitting of the \ion{Fe}{II}~615~nm line, \citet{2005A&A...443L...7B}
deduce that dark cores possess weaker fields than their surroundings.

\subsection{Faculae}
\label{sec:faculae}

\begin{figure*}[!ht]
\includegraphics[width=\textwidth]{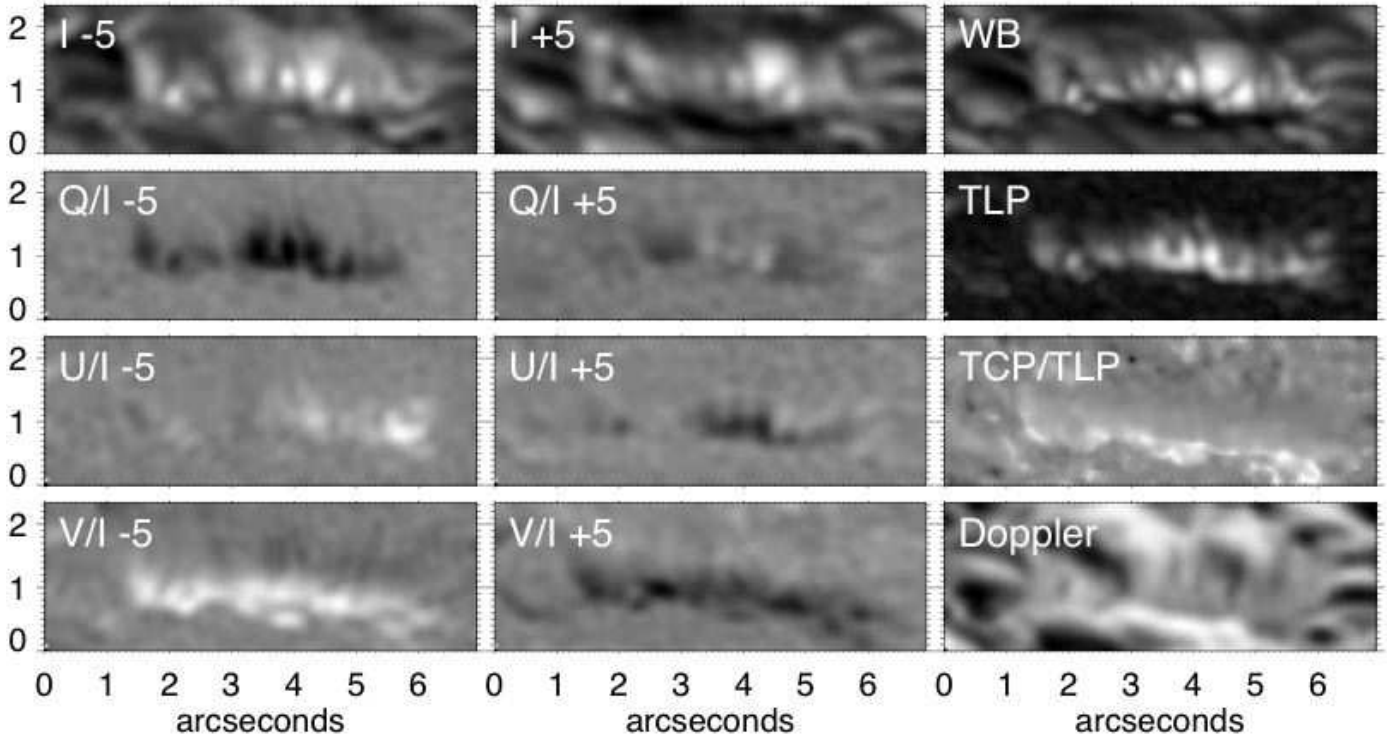}
\caption{Faculae observed on 7-Jan-2006. The left column shows Stokes
  maps at -5~pm, the middle column at +5~pm. The Stokes maps Q/I, U/I,
  V/I saturate at $\pm$10\%. The right column shows from top to
  bottom: wide-band (WB), total linear polarization (TLP), the ratio
  of total circular over linear polarization, and a map of the Doppler
  signal computed from the $\pm$5 Stokes I images. The limb direction
  is towards the top of the figures.}
\label{fig:faculae}
\end{figure*}

Another example of the spatial resolution and sensitivity that can be
achieved is provided by the facular region surrounding AR10845
observed on 7-Jan-2006. The active region is observed at a large
inclination angle ($\mu=0.39$), resulting in low contrast and few
structures with spatial scales at or below the diffraction limit. The
region shows several clusters of faculae surrounding the main sunspot.
Figure~\ref{fig:faculae} shows an overview of the different maps of
one such facular cluster.

The wideband image (``WB'') shows the faculae clearly identifiable as
a row of bright features, showing striations perpendicular to the
limbward direction and a sharp, dark lane on the disk-center side.
Such fine-structure in faculae is earlier reported and discussed from
high-resolution SST images by \citet{Lites:2004zr} and
\citet{2006ApJ...646.1405D}.
The faculae also clearly show up in the Stokes maps, with a similar
degree of fine structuring at the location of the faculae in the
circular and linear polarization maps as in Stokes I and wide-band.

In the -5~pm Stokes V/I map, we clearly see positive signal (white)
from magnetic fields pointing towards us at the disk-center side of
the faculae and negative signal (dark) from magnetic fields pointing
away from us on the limb side. A similar trend, with opposite sign, is
visible in the +5~pm V/I map, although the signals are slightly
weaker. This behavior is in agreement with the interpretation of
faculae as magnetic flux concentrations, fanning out horizontally over
its surroundings when the gas pressure falls below the magnetic
pressure.

The TLP map shows the presence of significant field in the plane
perpendicular to the LOS in the middle part of the faculae, but it
shows nearly no signal where the -5~pm V/I map shows the strongest
negative signal. This is probably the region where the magnetic field
is pointing only in our direction and is indicative of a
well-organized magnetic field structure as described above.

In all the polarization maps, a clear correlation can be seen between
the presence of striations, and the complete absence of any magnetic
signal. 
In the map with the ratio of circular over linear polarization
(TCP/TLP) at the location of the striations no variations can be
discerned. 
This indicates the decrease in polarization signal is not caused by a
variation of the magnetic field orientation but by a decrease in
magnetic field strength.
This is consistent with the explanation by \citet{2004ApJ...610L.137C}
that facular striations are caused by variations in the field strength
in the magnetic elements.

The numerical simulations by \citet{2004ApJ...607L..59K} and
\citet{2004ApJ...610L.137C} are successful in explaining the detailed
morphology of faculae observed at high spatial resolution.
The enhanced brightness of faculae are caused by the low density in
the atmosphere of magnetic elements as compared to the surrounding
medium. 
This allows the observer to look deeper into the hot wall of the
granule behind the element and causes excess brightness as compared to
neighboring rays going through non-magnetic atmospheres.
Horizontal variations in the magnetic field lead to associated
variations in density and opacity which explains the observation of
the striations in the direction perpendicular to the limbward
direction.
\citet{2006ApJ...646.1405D} give an extended discussion of the spatial
and temporal variation of faculae comparing time sequences of
high-resolution observations and numerical simulations. 

\citet{2007ApJ...661.1272B} observed reduced Stokes V/I signal in
facular striations using high resolution SST magnetograms. Here, we
extend these observations with measurements in linear polarization
which indicate that it is the decrease in the magnetic field that
cause striations and not variations in the magnetic field orientation.

The individual Q/I and U/I maps, show significantly different
behaviour on the red and the blue side of the line center, although
the line profile is naively expected to be symmetric. Notably, in the
red wing of the line the polarity of the Q/I and U/I signals is seen
to change sign repeatedly in the same row of faculae, whereas it does
not do so in the blue wing.

The 2-point Dopplermap shows that there is some variation in the
Doppler signal across the faculae. 
This indicates that the asymmetry between the red and blue wing maps
might be caused by velocities along the line of sight.  Without
additional spectral information available, it remains unclear what the
cause of this asymmetry is, or why it is particularly irregular in the
red side of the line.

\section{Discussion}
We have obtained near diffraction limited Full Stokes images of the
solar surface, with a S/N of $2\cdot 10^{-3}$ in Stokes V only mode
and $4\cdot 10^{-3}$ in Full Stokes mode, translating to a noise level of
$\approx 1.3\cdot 10^{-3}$ and $\approx 2.5\cdot 10^{-3}$ compared 
to the continuum intensity, a more conventional quantity used in 
spectropolarimetry. This translates to a maximum
sensitivity $33$~Mx~cm$^{-2}$ and $66$~Mx~cm$^{-2}$ respectively in
the weak field approximation. Although only 2 wavelengths were
observed, this data demonstrates that obtaining diffraction limited,
high S/N Full-Stokes images using 1\,m-class ground-based solar
telescopes is possible.

The observed features in penumbral dark cores and faculae are
consistent with recent result from numerical modeling and evidence
from Stokes V only data.

Despite the quality of the results, we note that there are at least
two major limitations in our current observing system that leave room
for significant improvements in terms of sensitivity and accuracy.

The SOUP filter is a Lyot filter, which means that it has a linear
polarizer that transmits only half of the incoming light. Due to the
large number of optical elements and the way in which it filters the
light, actual filter transmission is, at $\sim$15\%, 
well below that of any good quality modern filter, which may have 4--5
times higher transmission. With a prefilter transmission of
$\approx$50\%, the total transmission is only $\approx$7\%.
Moreover, the SOUP filter has a relatively long line position change time of
$\approx$6~s, implying that in our current scheme of $\approx$14~s acquisition
time per line position, 40\% of the available photon
collection time is lost by filter tuning.
A modern piezo-tuned filter system can typically tune in a matter of a few ms, 
so that it can change line position during the readout time of the camera. Not only
does this eliminate the time lost by filter tuning, an acquisition scheme,
better than the one used here, with the filter states interleaved as well as the
LCVR states, then becomes possible, with all the associated benefits in terms
of homogenizing the seeing conditions for all states.
Furthermore, by cycling continuously through a number of line
positions, one can decide a-posteriori what time scale is acceptable for the 
features to be studied and apply image restoration on a dataset covering the 
appropriate fraction of that feature's evolution time.

The readout time of the Sarnoff cameras is fixed, at 10~ms, so that
with an exposure time of 15~ms and a compulsory "dead" time of 2~ms
related to the chip size and the shutter speed, a total of 45\% of the
available photon collection time is lost. High-speed frame-transfer
cameras would therefore allow for another increase in efficiency of
approximately a factor 2, but reduce the possibility of interleaving
filter states, due to the finite tuning times of LCVRs and narrowband
filters.

The overall efficiency of the current setup can thus be estimated at a
mere 2-3\%, a figure that can surely be improved upon.

Separate from the efficiency aspect, there is the limitation from the
filter tuning time on the possible number of line positions. With a
tuning time of several seconds, the evolution time of the Sun does not
permit for more than a few wavelength points. An increased number of
line positions would, however, provide a highly desired enhanced
polarimetric accuracy as compared to the current two, as discussed in
detail by \cite{2002SoPh..208..211G}.

A Fabry-Perot Interferometer (see e.g., IBIS
\citep{2003MmSAI..74..796C,2003MmSAI..74..815R}, and TESOS
\citep{1998A&A...340..569K})
is a system that can effectively
decrease some of the shortcomings of the current setup. Recently, such an 
instrument has been installed at the SST (see \cite{2006A&A...447.1111S} for a 
design study), the transmission of which is 3--4 times higher than for the SOUP 
filter and in addition a polarizing beamsplitter is used
to ensure maximum use of the available light. Moreover, changing line
position can be accomplished in less than a camera readout time ($<10$ms),
allowing for multiple line positions to be covered in a fraction of a
second.  The overall performance increase in future observations is
therefore expected to yield significantly more accurate results than
was possible with the instruments used here.

\begin{acknowledgements}
The Swedish 1-m Solar Telescope is operated on the island of La Palma
by the Institute for Solar Physics of the Royal Swedish Academy of
Sciences in the Spanish Observatorio del Roque de los Muchachos of the
Instituto de Astrof{\'\i}sica de Canarias.
This 
research was supported through grants 146467/420
and 159137/V30 of The Research Council of Norway.
We thank Bruce Lites for his invaluable contributions to the
development of the polarimetry calibration procedure of the SST. 
This research has made use of NASA's Astrophysics Data System.
\end{acknowledgements}


\end{document}